\documentclass[aps,nofootinbib,twocolumn,preprintnumbers,superscriptaddress,nobibnotes,floatfix,longbibliography]{revtex4-1}

\pdfoutput=1
\usepackage{amsmath,amsfonts,amssymb,mathrsfs}
\usepackage{hyperref,cleveref}
\usepackage[utf8]{inputenc}
\usepackage{esint}

\usepackage{graphicx}
\usepackage{color}
\usepackage{ragged2e}
\usepackage{tikz}

\usepackage[loose]{units}

\newcommand{\beqn}{\begin{equation}}%
\newcommand{\eeqn}{\end{equation}}
\newcommand{\dd}{\mathrm{d}}
\newcommand{\gmn}{g_{\mu\nu}}
\newcommand{\fmn}{f_{\mu\nu}}
\newcommand{\mpl}{m_\mathrm{Pl}}
\newcommand{\mFP}{m_\mathrm{FP}}

\graphicspath{{figs/}}

\makeatletter

\def\hhref#1{\href{http://arxiv.org/abs/#1}{arXiv:#1}}
\usepackage{xstring} 
\newcommand{\hhrefq}[1]{\IfSubStr{#1}{:}{\href{http://inspirehep.net/search?ln=en&ln=en&p=#1&of=hb&action_search=Search&sf=&so=d&rm=&rg=25&sc=0}{InSpires:#1}}{\hhref{#1}}}

\def\art{\@ifnextchar[{\eart}{\oart}}
\def\eart[#1]#2#3#4#5#6{{\rm #2}, {\em #3 \bf #4} {\rm (#6) #5} ({\em #1})}
\def\article{\@ifnextchar[{\earticle}{\oarticle}}
\def\oarticle#1#2#3#4#5#6{{\rm #1}, {``#6''}, {\rm #2 #3 (#5) #4}}
\def\earticle[#1]#2#3#4#5#6#7{{\rm #2}, {``#7''}, {\rm #3 #4 (#6) #5}  [\hhrefq{#1}]}
\def\hepart[#1]#2{{\rm #2, \sl#1}}
\def\heparticle[#1]#2#3{#2, { ``#3''} [\hhrefq{#1}]}

\begin{document}

\title{Vainshtein Screening in Bimetric Cosmology}

\author{Marvin L\"uben}
\email{mlueben@mpp.mpg.de}
\affiliation{Max-Planck-Institut f\"ur Physik (Werner-Heisenberg-Institut),
 F\"ohringer Ring 6, 80805 Munich, Germany}

\author{Angnis Schmidt-May}
\email{angnissm@mpp.mpg.de}

\affiliation{Max-Planck-Institut f\"ur Physik (Werner-Heisenberg-Institut),
 F\"ohringer Ring 6, 80805 Munich, Germany}

\author{Juri Smirnov}
\email{smirnov.9@osu.edu}
\affiliation{Center for Cosmology and AstroParticle Physics (CCAPP), The Ohio State University, Columbus, OH 43210, USA}
\affiliation{Department of Physics, The Ohio State University, Columbus, OH 43210, USA}

\begin{flushright} 
\texttt{ MPP-2019-260 }
\end{flushright}

\begin{abstract}
We demonstrate that the early universe in bimetric theory
is screened by the Vainshtein mechanism on an FLRW background.
The spin-$2$ mass serves as the cosmological Vainshtein scale in this case.
This allows us to quantitatively address early universe cosmology.
In particular, in a global analysis, we study data from the cosmic microwave background
radiation and local measurements of
the Hubble flow. We show that bimetric cosmology resolves the discrepancy in the local and
early-time measurements of the Hubble scale via an effective phantom dark energy component.
\end{abstract}


\maketitle

\section{Introduction}

Modifications and extensions to the theory of general relativity (GR) are highly
restricted~\cite{Lovelock:1971yv,Lovelock:1972vz}.
One possible direction is the addition of new degrees of freedom to the massless spin-$2$ field
in a consistent manner.
Ghost-free bimetric theory
describes a massless and a massive spin-$2$ field with fully non-linear 
(self-)interactions~\cite{Pauli:1939xp,Fierz:1939ix,vanDam:1970vg,Zakharov:1970cc,Vainshtein:1972sx,Boulware:1973my,deRham:2010ik,deRham:2010kj,Hassan:2011vm,Hassan:2011hr,Hassan:2011tf,Hassan:2011ea,Hassan:2011zd,deRham:2014zqa,deRham:2014naa,deRham:2014fha}. 
It contains both general relativity and massive gravity in certain parameter limits in which
the massive or the massless field decouples, respectively. As such, bimetric theory
fills a gap in the list of consistent field theories for massive and massless particles with spin up to
$2$ and represents an extended gravitational theory with a rich phenomenology.

Bimetric theory can address several open questions in modern cosmology.
The theory gives rise to self-accelerating solutions where the interaction energy
between the massive and massless spin-$2$ field acts as dynamical dark
energy~\cite{DAmico:2011eto,vonStrauss:2011mq,Comelli:2011zm,Akrami:2012vf,Volkov:2011an,Konnig:2013gxa,DeFelice:2014nja,Konnig:2015lfa}.
The modification of the gravitational potential (as compared to GR) affects the required
Dark Matter abundance from galactic to galaxy cluster scales~\cite{Platscher:2018voh}.
Also, the massive spin-$2$ field itself serves as a candidate for Dark
Matter~\cite{Babichev:2016hir,Babichev:2016bxi,Chu:2017msm}.
Despite these successes, the stability of perturbations around the cosmological
background~\cite{Comelli:2012db,Konnig:2014dna,Konnig:2014xva,Lagos:2014lca,DeFelice:2014nja,Konnig:2015lfa,Akrami:2015qga}
still poses an open problem\footnote{Linear perturbation theory breaks down at a scale that can be pushed to arbitrary
early times~\cite{Akrami:2015qga}.
This is precisely the GR-limit of the theory.}.
At early times, certain perturbations become large, rendering linear perturbation
theory invalid.

In static systems with spherical symmetry an analogous behavior
occurs~\cite{Vainshtein:1972sx,deRham:2012fw,Babichev:2013usa,Babichev:2013pfa,Enander:2015kda}:
Non-linear effects in the perturbations are relevant and
render the linear approximation invalid. This is the well-known
Vainshtein screening mechanism~\cite{Vainshtein:1972sx} that restores
GR in spacetime regions where the energy density is large (compared to the spin-$2$ mass).
In massive gravity, this behavior is crucial because it cures the so-called 
vDVZ discontinuity~\cite{vanDam:1970vg,Zakharov:1970cc}.

In this paper, we give a physical argument for the cosmological version of the Vainshtein mechanism.
We find that the spin-$2$ mass sets the energy scale at which Vainshtein screening sets in.
Above this energy scale, the energy density coming from the interaction of the massive and massless mode is suppressed.
We show that the non-linear massless and massive modes indeed decouple in this regime.
It then becomes clear that the universe transitions from a Vainshtein screened period at early
times into a late time de Sitter phase.
We use this insight to discuss the instability of linear perturbations around the cosmological background in the light of existing results.

The Vainshtein mechanism is also known to play a crucial role in scalar-tensor galileon cosmologies~\cite{deRham:2011by}.
It has been pointed out, that at early times the galileon dynamics is governed by self-interactions, which results in a suppressed energy density relative to matter and radiation~\cite{Chow:2009fm}.
In this work we focus on an analogous effect in the framework of bimetric theory.

We use our result to address a highly debated problem of the
$\Lambda \rm CDM$ model, the $H_0$-tension.
As discussed in the literature~\cite{Riess:2016jrr,Riess:2018byc,Bonvin:2016crt,Birrer:2018vtm,Ade:2015xua,Aghanim:2018eyx,Efstathiou:2013via,Addison:2015wyg,Aghanim:2016sns,Aylor:2018drw,Mortsell:2018mfj,Aghanim:2016yuo,Follin:2017ljs,Dhawan:2017ywl}, CMB observations tend to
predict a lower Hubble scale than local measurements. We demonstrate that this
tension can be resolved in a minimal realization of bimetric theory with only two more parameters
than the $\Lambda \rm CDM$-model. These are the mass of the spin-$2$ field and
its coupling constant to ordinary matter. We perform a fit to the CMB,
Cepheid, supernova Type Ia (SN Ia) and baryon-accoustic-oscillations (BAOs)
observables and demonstrate that the tension
with the current data is alleviated. Instead, all observables are within
$1\sigma$ error intervals. This is achieved via an effective phantom phase of the
dark energy component in the redshift interval $1\lesssim z\lesssim 10$.
Finally, we comment on the compatibility between the different observables and 
argue that the $\Lambda\rm CDM$ and
the $\beta_0\beta_1\beta_4$ models are distinguishable with future measurements.

\section{Review of bimetric theory}
\label{sec:intro-bimetric-theory}

In this section, we review some technical aspects of the bimetric theory.
Readers familiar with bimetric theory might want to jump
directly to the next section. For a review on bimetric theory
we refer to Ref.~\cite{Schmidt-May:2015vnx}.
For an introduction to the broad field of massive gravity we refer to Refs.~\cite{Hinterbichler:2011tt,deRham:2014zqa}.

\subsection{Action and equations of motion}

The action of bimetric theory with standard model matter minimally coupled to one metric is~\cite{deRham:2010ik,deRham:2010kj,Hassan:2011zd}
\begin{flalign}\label{eq:bim-action}
    S = & \frac{m_g^2}{2}\int\text{d}^4 x \left(\sqrt{-g}\,R(g)+\alpha^2 \sqrt{-f}\,R(f)\right)\nonumber\\
         -& m_g^2 \int\text{d}^4 x \sqrt{-g}\, V(g,f) + \int\text{d}^4 x \sqrt{-g}\,\mathcal L_m(g,\Phi)\,,
\end{flalign}
where $R(g)$ and $R(f)$ are the Ricci scalars of the metric tensors $\gmn$ and $\fmn$, respectively.
The parameter $m_g$ is the Planck mass of $\gmn$ and $\alpha$ parametrizes the ratio to the
Planck mass of $\fmn$.
The interaction part of the Lagrangian has been derived prior to the bimetric formulation in~\cite{deRham:2010ik,deRham:2010kj}.
The potential, whose structure is entirely fixed by the absence of ghost instabilities, reads
\begin{flalign}
    V(g,f)=\sum_{n=0}^4 \beta_n e_n\left(\sqrt{g^{-1}f}\right)
\end{flalign}
in terms of the elementary symmetric polynomials $e_n$. 
The interaction parameters $\beta_n$ have mass-dimension $2$ in our parametrization.
Matter fields, collectively denoted by $\Phi$, minimally couple to the metric $\gmn$ via the
generic matter Lagrangian $\mathcal L_{\rm m}$.

Varying the action~\eqref{eq:bim-action} with respect to the metric tensors yields two
sets of modified Einstein equations,
\begin{equation}
    G^g_{\mu\nu}+V^g_{\mu\nu}=\frac{1}{m_g^2} T_{\mu\nu}\,,\quad
    G^f_{\mu\nu}+V^f_{\mu\nu}=0\,,
\end{equation}
where $G^g_{\mu\nu}$ and $G^f_{\mu\nu}$ are the Einstein tensors of $\gmn$ and
$\fmn$, respectively. The variations of the potential yield
\begin{align}
    V^g_{\mu\nu} &= g_{\mu\lambda}\sum_{n=0}^3 (-1)^n\beta_n Y^\lambda_{(n)\nu}\left(\sqrt{g^{-1}f}\right)\,,\\
    V^f_{\mu\nu} &= f_{\mu\lambda}\sum_{n=0}^3 (-1)^n\beta_{4-n} Y^\lambda_{(n)\nu}\left(\sqrt{f^{-1}g}\right)\,,
\end{align}
where the matrices $Y^\lambda_{(n)\nu}(S)$ are sums of different powers and contractions of
the matrix $\sqrt{g^{-1}f}$~\cite{Hassan:2011vm}.
Since matter minimally couples to $\gmn$, the stress-energy tensor for $\gmn$ is given by
\begin{equation}
    T_{\mu\nu}=\frac{-2}{\sqrt{-g}}\frac{\delta \sqrt{-g} \mathcal L_m}{\delta g^{\mu\nu}}\,.
\end{equation}
For usual matter (invariant under diffeomorphisms), stress-energy is conserved
\begin{equation}\label{eq:conservation}
    \nabla_\mu T^{\mu\nu}=0\,,
\end{equation}
where $\nabla_\mu$ is the covariant derivative compatible with $\gmn$.
Due to the Bianchi identity $\nabla^\mu G^g_{\mu\nu}=0$, this implies an on-shell
constraint on the potential,
\begin{equation}
    \nabla^\mu V^g_{\mu\nu}=0\,,
\end{equation}
referred to as Bianchi constraint.

\subsection{Proportional background}

The theory has a well-defined mass spectrum only around proportional backgrounds
where the two metrics are related as $\fmn=c^2 \gmn$ for some non-vanishing constant
$c$~\cite{Hassan:2011zd,Hassan:2012wr}.  The solution is an Einstein
background, $R_{\mu\nu}(f)=R_{\mu\nu}(g)=\Lambda\gmn$,
with a cosmological constant given by
\begin{align}\label{eq:prop-bkg-consistency}
\begin{split}
    \Lambda = & \beta_0 + 3\beta_1 c + 3\beta_2 c^2 + \beta_3 c^3\\
    = & \frac{1}{\alpha^2}\left( \frac{\beta_1}{c} + 3\beta_2 + 3\beta_3 c +\beta_4 c^2 \right)\,.
    \end{split}
\end{align}
The equality of the first and second line is required for the consistency of the vacuum solution. 
It is a quartic polynomial in $c$ and each root corresponds to a different Einstein solution to the bimetric field equations. 

The Fierz-Pauli mass of the massive spin-2 mode that propagates on the proportional 
background is given by
\begin{equation}\label{eq:FP-mass}
    \mFP^2 = \left(1+\frac{1}{\alpha^2 c^2}\right)c (\beta_1 + 2 \beta_2 c + \beta_3 c ^2)\,,
\end{equation}
in terms of the bimetric parameters. For later use, let us introduce the short-hand notation $\bar\alpha=\alpha c$.
In the rest of the paper, we will refer to $\bar\alpha$, $\mFP$, and $\Lambda$ as
the physical parameters of a solution~\cite{Luben:2020xll}.

\subsection{FLRW solutions}

To describe the universe on large scales, we assume spacetime to be homogeneous and isotropic
according to the cosmological principle. 
Both metrics take on FLRW form~\cite{Volkov:2011an,vonStrauss:2011mq,Comelli:2011zm},
\begin{align}
\begin{split}\label{eq:FLRW-metrics}
    &\text{d}s^2_g = a(\eta)^2\left(-\text{d}\eta^2 +\text{d}\vec x^2  \right)\\
    &\text{d}s^2_f = b(\eta)^2\left(-(1+\mu(\eta))^2\text{d}\eta^2 +\text{d}\vec x^2  \right)\,,
\end{split}
\end{align}
where we fixed the time-reparametrization invariance such that we work in conformal time $\eta$ of the $g$-metric. 
The scale factors $a(\eta)$ and $b(\eta)$ of the metrics $\gmn$ and $\fmn$  are functions of time only. 
The $f$-metric lapse $(1+\mu)$ parameterizes the relative twist between the coordinate systems of the two metrics. In this sense, it is similar to a St{\"u}ckelberg field since it would be shifted by 
time reparametrizations of the metric $\fmn$.
From now on, we suppress the $\eta$-dependence. 
For convenience, let us introduce the Hubble rate and the 
ratio of the scale factors as
\begin{flalign}
    \mathcal H = \frac{\dot a}{a} \,,\quad y=\frac{b}{a}\,,
\end{flalign}
where the dot represents derivative w.r.t.~conformal time. 
The conformal and physical Hubble rates are related via $\mathcal H = aH$.

We assume the universe to be filled with a perfect fluid with stress-energy tensor,
\begin{flalign}
    T^{\mu\nu} = (\rho_{\rm m}+p_{\rm m})u^\mu u^\nu + p_{\rm m} g^{\mu\nu}\,,
\end{flalign}
where $\rho_{\rm m}$ is the energy density and $p_{\rm m}$ the pressure of the matter fluid.
They are related via the linear equation of state $w_{\rm m}=p_{\rm m}/\rho_{\rm m}$. $u^\mu$ 
is the $4$-velocity of the fluid. The 
conservation \cref{eq:conservation} leads to the continuity equation,
\begin{flalign}\label{eq:continuity}
    \dot\rho_{\rm m} + 3\mathcal{H}(1+w_m) \rho_{\rm m} =0\,,
\end{flalign}
which is solved by
$\rho_{\rm m}=\rho_\mathrm{m,0} a^{-3(1+w_m)}$, 
where $\rho_\mathrm{m,0}$ is an integration constant. The physical
scale factor is related to the redshift as $a=(1+z)^{-1}$.

The Bianchi constraint on the dynamical branch\footnote{Another solution to the Bianchi 
constraint is the algebraic branch, where the ratio of the scale factors is fixed, $y=\text{const}$. 
This branch was studied in the literature and found to be pathological, 
see, e.g.,~Refs.~\cite{Comelli:2012db,Cusin:2015tmf,Konnig:2015lfa}.}
is solved by
\begin{flalign}\label{eq:Stuckelberg-definition}
    \mu = \frac{\dot y}{\mathcal H y} = \frac{y^\prime}{y}\,,
\end{flalign}
where we have introduced the derivative w.r.t.~$e$-folds, 
${}^\prime=\text{d}/\text{d}\ln a$. 
On the dynamical branch, the modified Friedmann equations for
$\gmn$ and $\fmn$ read,
\begin{flalign}\label{eq:mod-friedmann}
    3\mathcal H^2 & = a^2\left(\frac{\rho_{\rm DE}}{m_g^2}+\frac{\rho_{\rm m}}{m_g^2}\right)\,,\\
    3\alpha^2 \mathcal H^2 & = a^2\left( \frac{\beta_1}{y} + 3\beta_2 + 3\beta_3 y + \beta_4 y^2 \right)\,.
\end{flalign}
Here, we defined the energy density coming from the potential as
\begin{equation}\label{eq:dark-energy-density}
	\frac{\rho_{\rm DE}}{m_g^2} = \beta_0 + 3\beta_1 y + 3\beta_2 y^2 + \beta_3 y^3\,,
\end{equation}
which can be interpreted as dynamical dark energy.
Combining both Friedmann equations yields a quartic polynomial for $y$,
\begin{flalign}\label{eq:quartic-pol-y}
    \alpha^2\beta_3 y^4+(3\alpha^2\beta_2-\beta_4) y^3+3(\alpha^2\beta_1-\beta_3)y^2 & \nonumber\\
    +\left(\alpha^2\beta_0 -3\beta_2 + \frac{\alpha^2\rho_{\rm m}}{m_g^2}\right)y -\beta_1 & = 0\,,
\end{flalign}
that can be thought of determining $y$ as a function of $\rho_{\rm m}$. 
The polynomial has in general up to four real-valued roots, of which only one 
(the so-called finite branch) is physical~\cite{Konnig:2013gxa,Konnig:2015lfa}. 
On this branch, the ratio of scale factors evolves from $y=0$ in the early universe 
(where $\rho_{\rm m}=\infty$), to a finite constant value $y=c$ in the asymptotic
future (where $\rho_{\rm m}=0$, i.e. de Sitter space). 

Taking the derivative w.r.t.~$e$-folds and using the conservation~\cref{eq:continuity}, 
we can express $y'$ as a function of $y$ as~\cite{Akrami:2012vf}
\begin{flalign}\label{eq:y-prime}
    y^\prime=\frac{3(1+\omega_{\rm m})\alpha^2 y^2 \rho_{\rm m}/m_g^2}{\beta_1-3\beta_3 y^2-2\beta_4 y^3 + 3\alpha^2 y^2 (\beta_1+2\beta_2 y + \beta_3 y^2)}\,,
\end{flalign}
where $\rho_{\rm m}$ is a function of $y$ via~\cref{eq:quartic-pol-y}.
In terms of the dynamical mass parameter
\begin{flalign}
    m_\mathrm{eff}^2=\frac{1+\alpha^2y^2}{\alpha^2y^2}y(\beta_1+2\beta_2 y+\beta_3 y^2)\,,
\end{flalign}
we can rewrite \cref{eq:y-prime} as
\begin{flalign}
    \frac{y'}{y}=\frac{(1+w_{\rm m})\rho_{\rm m}/m_g^2}{m_\mathrm{eff}^2-2H^2}\,.
\end{flalign}
In the following discussion we always assume an expansion history on the finite branch\footnote{The
finite
branch is only well-defined for $\beta_1>0$
~\cite{Konnig:2013gxa}.
Hence, in this paper we exclude models with $\beta_1=0$.}
implying $m_\mathrm{eff}^2>2H^2$~\cite{Fasiello:2013woa}.

\section{Vainshtein screening}

The Vainshtein mechanism was discovered in systems, where due to a locally increasing
gravitational field, GR is restored by non-linear interactions~\cite{Vainshtein:1972sx,Babichev:2013usa}.
We translate the analysis to the case where the gravitational field varies in time.
By studying the cosmological background and linear perturbations we identify
the energy scale at which non-linearities become important. Due to our analogy
this can be interpreted as a cosmological Vainshtein mechanism. 

\subsection{The standard Vainshtein mechanism}
On the technical level, the Vainshtein screening effect is caused by the strong coupling of the longitudinal helicity-$0$ mode of the massive spin-$2$ field. 

In the case of a point source in bimetric theory, 
the critical radius below which the non-linearities of the fields become strong enough 
is given by~\cite{Babichev:2013pfa}, 
\beqn 
r_{\rm V} = \left(\frac{r_{\rm S}}{m_\mathrm{FP}^2}\right)^{1/3}\,.
\eeqn 
Here, the Schwarzschild radius for a source of mass $M$ is given by
\beqn
r_{\rm S} =  \frac{M}{4 \pi \, m_g^2}\,.
\eeqn

For a central, point-like source (or on scales much larger than the
massive object itself), the induced gravitational potential 
is then given by~\cite{Babichev:2013pfa,Enander:2013kza,Platscher:2016adw,Babichev:2016bxi},
\begin{flalign}\label{eq:gravitational-potential}
    \phi(r) =
    \begin{cases}
        - \frac{1}{m_g^2} \frac{1}{r}   & r \ll r_{\rm V}\,, \\
         - \frac{1}{\mpl^2} \left( \frac{1}{r} + \frac{4\bar\alpha^2}{3}\frac{e^{ -m_\text{FP} r}}{r}\right)  & r \gg r_{\rm V} \,.
    \end{cases}
\end{flalign}
Inside the Vainshtein sphere, i.e. when $r\ll r_{\rm V}$, the gravitational potential is the
same as in GR with Planck mass $m_g$. Outside the Vainshtein sphere, the
massive spin-$2$ field propagates, contributing an attractive Yukawa term to the potential.
On scales much larger than the Compton wavelength, $r\gg \mFP^{-1}$,
the Yukawa term is suppressed and the potential coincides with the one in GR,
however with a different Planck mass given by
$\mpl=\sqrt{1+\bar\alpha^2}\, m_g$, compared to the one inside the Vainshtein sphere.
In this sense, there are two different scales on 
which GR is recovered, but with different Planck masses.
Note that we neglect an asymptotic cosmological constant in~\cref{eq:gravitational-potential}.

Given a Schwarzschild geometry, on a technical level, the crucial indicator for the transition from the Vainshtein screened regime to the massive gravity regime, is the radial, relative metric twist $\mu(r)$~\cite{Babichev:2013usa,Platscher:2016adw}. This function drops off quickly outside the Vainshtein regime and can be used as a small parameter in perturbation theory. Inside the Vainshtein regime $\mu(r)$ becomes large and the perturbative expansion breaks down. On the non-linear level, however, in this regime, the GR solution is recovered.
We will show in the following, that the same behavior is present on the FLRW
background and use the temporal metric twist function $\mu(\eta)$, introduced above,
to study when it becomes non-linear.

For a source of finite size, the Vainshtein mechanism can be at work, 
once the radius within which the matter is concentrated is smaller than the Vainshtein radius itself. 
This indicates that there exists a minimal density for the Vainshtein mechanism. 
In the next section, we will elaborate on this observation and apply its logic to cosmology.

\subsection{The Vainshtein mechanism in cosmology}\label{sec:Vainshtein-cosmo}

In this section, we will explore possible implications of the standard Vainshtein mechanism 
for the Universe filled with a homogeneous energy density.
From analogy with the galileon cosmology~\cite{Chow:2009fm, deRham:2011by}, we expect that a cosmological Vainshtein regime would be present in bimetric theory, and restore a GR-like solution at early times and large densities in the universe.

In particular, using properties of the solution for a static, 
spherically symmetric system around a compact source, we would like to answer the following question: 
What is the critical density $\rho_\mathrm{c}$
of a homogeneous mass distribution, for which the entire mass lies inside its own Vainshtein radius?
We can only give a very rough estimate for this value since the standard expression for the Vainshtein
radius is derived assuming that its value is larger than the radius of the source. 
In the following, we simply extend its definition to all possible configurations.

Let us thus consider a spherical, constant mass or energy distribution $\rho(r)=\,$const. 
This distribution can be infinitely extended with
radius $R=\infty$ or have a large but finite radius $R$;
for our argument below, this will make no difference.
The mass $M(r)$ enclosed within a distance $r<R$ from the center of the mass 
distribution is,
\beqn
	M(r)=\int_0^r\dd^3r'~\rho(r')=\rho \int_0^r\dd^3r'= \frac{4\pi}{3}r^3\rho\,.
\eeqn
The Schwarzschild radius corresponding to this enclosed mass is,
\beqn
	r_\mathrm{S}=\frac{ M(r)}{4 \pi \, m_g^2}=\frac{r^3\rho }{3\, m_g^2}\,.
\eeqn
The Vainshtein radius corresponding to the mass $M(r)$ is,
\beqn\label{vsr}
	r_{\rm V}=\left(\frac{r_\mathrm{S}}{m_\mathrm{FP}^2}\right)^{\frac{1}{3}}
=\left(\frac{ \rho}{3m_g^2m_\mathrm{FP}^2}\right)^{\frac{1}{3}}r\,.
\eeqn
The mass within the radius $r$ fits precisely inside its own Vainshtein radius,
when $r=r{\rm V}$. This gives the value for the critical density,
\beqn\label{critrho}
\rho_\mathrm{c}= 3m_g^2m_\mathrm{FP}^2 \,.
\eeqn
For less dense systems with $\rho<\rho_\mathrm{c}$, the Vainshtein radius is smaller
than the radius that encloses its corresponding mass. 
For denser systems with $\rho>\rho_\mathrm{c}$, the mass enclosed by a radius $r$ lies entirely inside
its own Vainshtein radius.

\begin{figure}
\centering
\includegraphics[scale=0.7]{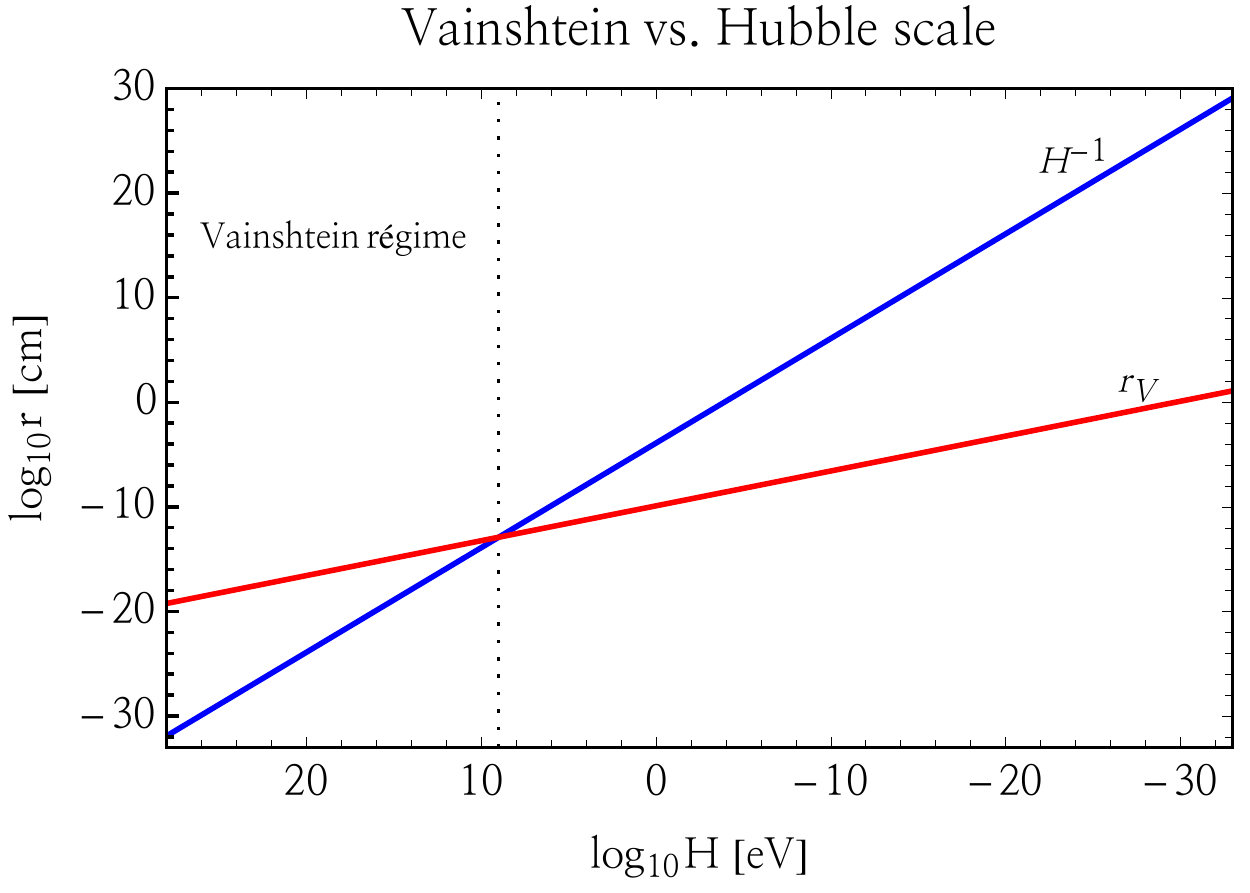}
\caption{\raggedright
 For a spin-2 mass of $\mFP=1\,\unit{GeV}$, 
 the size $H^{-1}$ of the Universe and its cosmological Vainshtein radius $r_{\rm V}$
are plotted as functions of the Hubble rate $H$. 
Since the universe expands faster than its Vainshtein sphere, it unavoidably becomes larger than this sphere.
That happens at the critical Hubble rate $H_\text{c}=\mFP$.}
\label{fig:Cosmologcial-vainshtein}
\end{figure}

We can now apply this result to cosmology, assuming that time evolution will not invalidate the arguments. 
Our reasoning suggests that there exists a critical value $\rho_\mathrm{c}$ for the 
homogeneous energy density $\rho(t)$ in the Universe, given by (\ref{critrho}).
Since $\rho(t)$ is time-dependent and increases with redshift, there will be a point in time when it passes
the value $\rho_\mathrm{c}$. Using Friedmann's equation, $H^2=\rho/(3m_g^2)$, with 
$\rho$ being the total energy density of the universe, this translates into
a critical value for the Hubble rate,
\beqn\label{hcrit}
H_\mathrm{c}= m_\mathrm{FP}\,.
\eeqn
Moving backward in time, we reach the point $t_\mathrm{c}$ at which
$H(t)$ passes the value $H_\mathrm{c}$.
At this time, the energy density takes the value $\rho_\mathrm{c}$, for which its corresponding mass lies 
entirely inside its own Vainshtein radius.

For a homogenous matter distribution like in cosmology, the Vainshtein radius scales linearly
with radius,  $r_{\rm V}\sim r$. 
Therefore, if the energy density of the universe is below
its critical value, $\rho<\rho_c$ (or equivalently $H< m_\mathrm{FP}$), each Hubble patch
is outside the Vainshtein regime. Analogously, for larger energy densities,
$\rho>\rho_c$, and hence at early times, each Hubble patch lies
within its Vainshtein region. Even though we apply a concept derived in a
spherically symmetric setup to a homogenous
energy distribution, the equations do not single out a preferred point in space.
This is consistent with translational invariance
of FLRW; our analogy does not spoil the cosmological principle.

Instead, this allows for another, rather curious interpretation. In the expanding universe, let us treat the Big Bang singularity as a central source
and measure the distance to it by the inverse Hubble scale $H^{-1}$.
From this perspective, the profile of the energy density distribution is
$\rho\sim H^2 \sim r^{-2}$ according to Friedmann's equation.
The corresponding Schwarzschild radius
is $r_{\rm S}=H^{-1}$.
Calculating the Vainshtein radius in this setup yields
\begin{flalign}
    r_{\rm V} = \left(\mFP^2 H\right)^{-1/3} \,,
\end{flalign}
which of course leads to the same critical Hubble rate $H_c$ derived above as
this is just the special case of the above analysis for $r=H^{-1}$. 

In \cref{fig:Cosmologcial-vainshtein}, we show the different scalings of the Vainshtein radius and of the size of the
observable universe as functions of the Hubble rate $H$. The cosmological Vainshtein radius scales as $r_{\rm V} \sim H^{-1/3}$ while the size
of the observable universe scales as $H^{-1}$. Therefore, at early times
when the Hubble rate is large, the Vainshtein radius is larger than the size of a
Hubble patch. At the critical value set by the spin-$2$ mass, the former surpasses the latter, and at late times a Hubble patch
is larger than its Vainshtein sphere.

In \cref{fig:bim-expansion-history}, the evolution of the universe is schematically depicted, and can be interpreted as follows. 
The Big-Bang singularity serves as an analog of the central source in spherically symmetric systems. 
Moving away from the source, i.e.~letting time pass, 
the evolution is expected to be governed by equations of motion equivalent to GR because we are inside
the Vainshtein sphere. When the Hubble rate becomes comparable to the critical value, 
the universe exits the Vainshtein regime and enters a transition phase
where the evolution is governed by the bimetric field equations. 
Asymptotically in the future, the universe approaches de Sitter space.
In this sense, also at late times, GR is effectively recovered.

We emphasize again that we made simplifying assumptions about the validity
of the expression for the Vainshtein radius\footnote{Note that the standard derivation
of the Vainshtein mechanism is restricted to scales $r$
which are much smaller than the Compton wavelength of the massive spin-2 field, $r\ll \mFP^{-1}$. 
When considering the entire Universe, the Vainshtein radius is $r_{V,\mathrm{crit}}=\mFP^{-1}$
and thus violates this condition.}.
Thus our arguments here can at most give very rough
estimates for the critical values of cosmological quantities. Nevertheless, these approximate values are supported by results in cosmological perturbation theory,
see \cref{sec:perturbations}.
Moreover, we observe interesting behavior of the solutions already at the background level,
as we demonstrate in the next section.

\begin{figure}
\centering
\begin{tikzpicture}
    \filldraw[fill=yellow] (0,0) circle [radius=0.2];
    \node at (0.5,0.2){$H^{-1}$};
    \draw (0,0) circle [radius=0.9];
    \draw [->] (0,0) -- (0.17,0.1);
    \node at (0,-0.3){};
    \filldraw (0,0) circle (2pt);
    \draw [->] (0,0) -- (-0.55,0.7);
    \node at (-0.7,0.8){$r_{\rm V}$};
    \filldraw[fill=yellow] (5,0) circle [radius=2.2];
    \draw (5,0) circle [radius=1.5];
    \node at (6.4,1.2){$H^{-1}$};
    \draw [->] (5,0) -- (6.83,1.2);
    \node at (5,-0.3){};
    \filldraw (5,0) circle (2pt);
    \draw [->] (5,0) -- (4.12,1.2);
    \node at (4.2,0.7){$r_{\rm V}$};
    \node at (0,-2.5){$H^{-1}\ll r_{\rm V}$};
    \node at (5,-2.5){$H^{-1}\gg r_{\rm V}$};
    \node at (0,2.5){Vainshtein (early times)};
    \node at (5,2.5){de Sitter (late times)};
\end{tikzpicture}
\caption{\raggedright 
Schematical depiction of the expansion of the universe. 
Since the Universe expands faster than its Vainshtein radius, 
it leaves the Vainshtein sphere at the critical Hubble rate.}
\label{fig:bim-expansion-history}
\end{figure}

\section{The spin-$2$ mass in cosmology}

In the previous section we have identified $H_{\rm c} = \mFP$ as the critical Hubble rate, at which the universe transitions from a Vainshtein screened to an unscreened phase.
In this section, we analyze how cosmological solutions to bimetric theory behave in the two regimes.
In order to present explicit results, we use the $\beta_0\beta_1\beta_4$-model as toy model for which we provide details in~\cref{append:b0b1b4}.

\subsection{Vainshtein screening of background cosmology}

In this section, we demonstrate at the level of Friedmann's~\cref{eq:mod-friedmann}
that the background cosmology indeed transitions from the GR 
to the bimetric phase exactly at the energy scale $\mFP$.

We start by discussing the scale factor ratio $y$.
As discussed in~\cref{sec:intro-bimetric-theory}, on the finite branch the scale factor ratio evolves from zero at early times to a constant determined by~\cref{eq:prop-bkg-consistency} in the asymptotic future.
More precisely, we can expand~\cref{eq:quartic-pol-y} for $\rho_{\rm m}/m_g^2\gg\beta_n$ to find
\begin{equation}
	\bar y = \bar\beta_1\left(\frac{ \rho_{\rm m}}{m_g^2}\right)^{-1} + \mathcal O\left(\frac{\rho_{\rm m}}{m_g^2}\right)^{-2}\,.
\end{equation}
Here we introduced the parameters $\bar y = \alpha y$ and $\bar \beta_n = \alpha^{-n}\beta_n$ for convenience~\cite{Luben:2020xll}.
Plugging this result into the expression for the dark energy density~\eqref{eq:dark-energy-density} yields
\begin{equation}
	\frac{\rho_{\rm DE}}{m_g^2} = \beta_0 + \frac{3\bar\beta_1^2 m_g^2}{\rho_{\rm m}} + \mathcal O\left(\frac{\rho_{\rm m}}{m_g^2}\right)^{-2}\,.
\end{equation}
Since $\beta_0$ is a constant we thus find that $\rho_{\rm m} \gg \rho_{\rm DE}$ in the early universe, as pointed out already in~\cite{vonStrauss:2011mq}.
Hence, the energy density coming from the interaction between the spin-2 fields does not contribute to the Hubble rate at early times.
This effect represents cosmological screening in close analogy to galileon cosmology~\cite{Chow:2009fm}.

Next, we want to study when this screening sets in.
As such, we determine the energy scale at which $\bar y'$ is at its maximum.
We therefore solve the equation $\bar y''=0$ and infer the corresponding Hubble rate.
However, analytical solutions could not be found for this polynomial in the
$\beta_0\beta_1\beta_4$-model due to its high degree.
Instead, we solve $\bar y''=0$ for the two submodels with $\beta_0=0$ and $\beta_4=0$, respectively.
The results in both cases are lengthy, but in the limit $\bar\alpha\ll1$ we get the remarkably simple result
\begin{align}\label{eq:critical-hubble-cosmo}
    H_* \simeq \mFP \,,
\end{align}
as the critical Hubble rate at which $\bar y''=0$ for both submodels, up to an $\mathcal O(1)$ factor\footnote{Note
that the limit $\bar\alpha\ll1$ has very different meaning in both submodels~\cite{Luben:2020xll}.
Since we fix one of the interaction parameters, one of the physical parameters is not free, but depends on
the other parameters. For the submodel where $\beta_0=0$ we get
$\bar\alpha^2=\Lambda/(3\mFP^2-\Lambda)$ while for the submodel with
$\beta_4=0$ the relation is
$\bar\alpha^2=\mFP^2/\Lambda-1$, cf.~\cref{eq:b0-phys-parameter-b0b1b4}.
Hence our limit in the former case
implies $\mFP^2\gg\Lambda$, while in the latter it
implies $\mFP^2\simeq\Lambda$. Consequently, for the $\beta_0\beta_1$-model
there is only one energy scale involved.
Expanding around $\bar\alpha\gg1$ (i.e. $\mFP^2\gg\Lambda$) instead
results in $H_*\simeq(\mFP\Lambda)^{1/3}$ as critical Hubble rate at which $y''=0$.
Only the $\beta_0\beta_1$-model gives rise to this behavior.}.

Let us provide a physical interpretation for the parameter $\bar y=\alpha y$.
The non-linear massless field $G_{\mu\nu}$ is given by~\cite{Hassan:2012wr}
\begin{equation}
	G_{\mu\nu} = g_{\mu\nu} + \alpha^2 f_{\mu\nu}\,,
\end{equation}
while the non-linear massive field is not unique.
On FLRW~\eqref{eq:FLRW-metrics}, the spatial components read $G_{ij} = a^2(1+\bar y^2)\delta_{ij}$.
Hence, in the limit $\bar y\ll 1$ the non-linear massless mode and the physical metric are aligned.
This implies that the non-linear massless and massive modes decouple.
This observation is consistent with the decoupling of the linear massless and massive mode in the limit $\bar\alpha\ll 1$~\cite{Akrami:2015qga,Luben:2018ekw}.
We conclude that $\bar y$ parametrizes the mixing of the massless and massive field on an FLRW background.

However, we also have to consider the temporal component of the non-linear massless field that is given by $G_{00}=-a^2(1+\bar y^2(1+\mu)^2)$.
The decoupling is not only controlled by $\bar y$, but also the St{\"u}ckelberg field $\mu$.
Hence, let us study the evolution of $\mu$.
At late times, the matter-energy density $\rho_{\rm m}$ vanishes implying
that $y'$ vanishes and the St{\"u}ckelberg field is small,
\begin{flalign}
    \mu \longrightarrow 0 \quad&\text{for}~~\eta\rightarrow +\infty\,,
\end{flalign}
cf.~\cref{eq:Stuckelberg-definition}.
At early times $\bar y$ approaches zero and the energy density diverges as $\rho_{\rm m}/m_g^2\sim \bar\beta_1/\bar y$.
The St{\"u}ckelberg field approaches the constant value,
\begin{flalign}
    \mu \longrightarrow3(1+w_{\rm m}) \quad&\text{for}~~\eta\rightarrow -\infty\,,
\end{flalign}
which is $\mathcal O(1)$.
The St{\"u}ckelberg field $\mu$ hence transitions from a large value at early times to a small value in the asymptotic future.
Although this behavior does not spoil the decoupling, 
it is analogous to the effect in the Schwarzschild geometry.
Here, $\mu$ asymptotes to a constant value when approaching the source~\cite{Babichev:2013pfa}.

\begin{figure}
\centering
\includegraphics[scale=0.65]{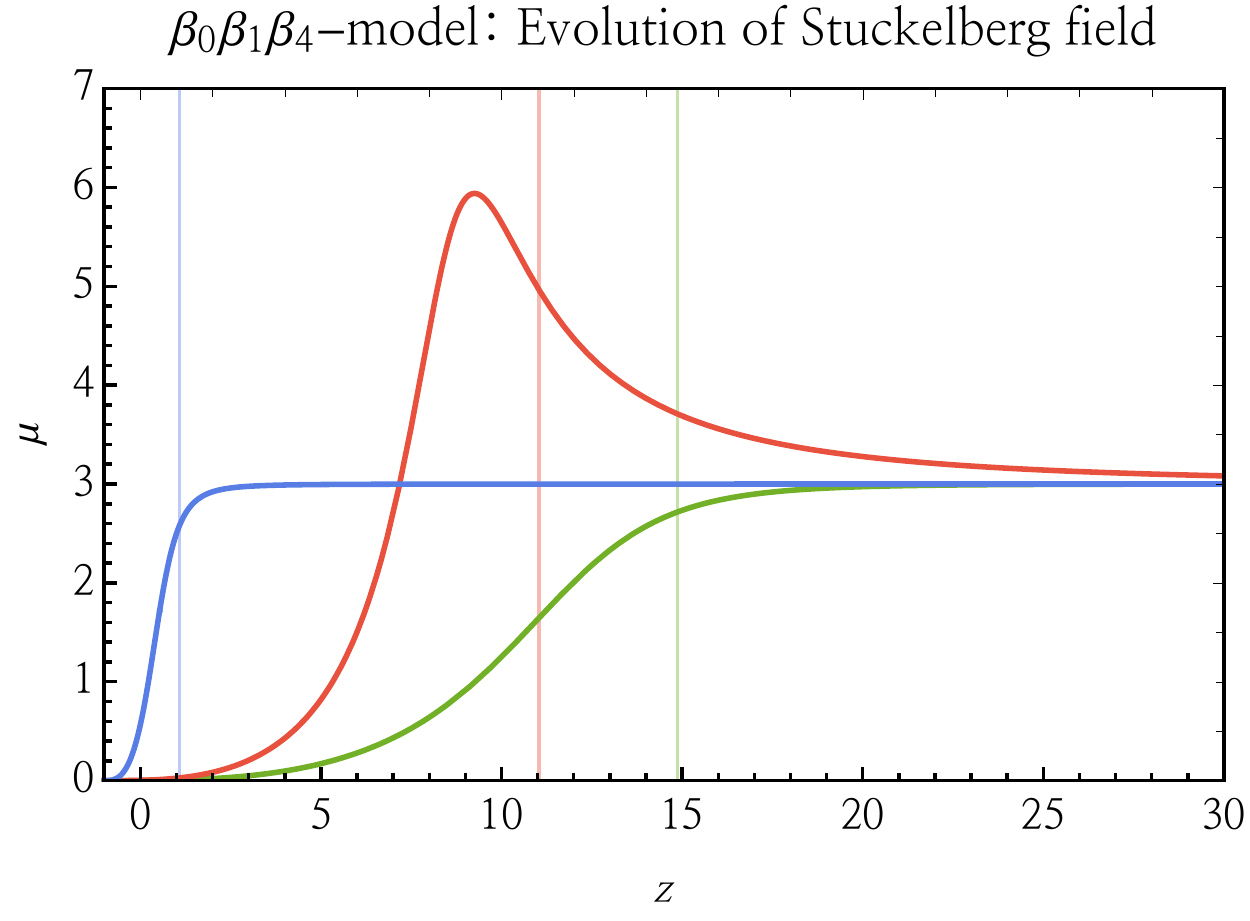}
\caption{\raggedright The time evolution of the St{\"u}ckelberg field
$\mu$ as a function of redshift $z$. For all lines we set
$\Lambda=0.7H_0^2$ and $\rho_\mathrm{m,0}/m_g^2=0.3H_0^2$ as exemplary values.
The bimetric parameters are $\mFP=H_0$ and $\bar\alpha=0.5$ (blue line),
$\mFP^2=10H_0^2$ and $\bar\alpha=2$ (red line), and
$\mFP^2=20H_0^2$ and $\bar\alpha=0.01$ (green line).
The vertical lines represent the redshift at which $H=\mFP$, respectively
In each case, $\mu$ transitions from the early time asymptotic value to
$0$ around the scale $\mFP$.}
\label{fig:b0b1b4-stuckelberg}
\end{figure}

In~\cref{fig:b0b1b4-stuckelberg}, we plot the St{\"u}ckelberg field
as a function of redshift $z$ in the $\beta_0\beta_1\beta_4$-model for three exemplary
cases with different mixing angles $\bar\alpha$ and spin-$2$ masses $\mFP$
as displayed in the caption.
The vertical lines represent the critical redshift at which $H=\mFP$.
Qualitatively we find that the St{\"u}ckelberg field $\mu$ indeed starts to deviate from its asymptotic value as soon as the Hubble rate is of the order of $\mFP$.
We have explicitly checked this for various
examples and find that the behavior is completely generic\footnote{Note that for
the case represented by the red line $\mu$ develops a peak. This happens only for models
with $\beta_0<0$ as we checked explicitly. For submodels with $\beta_0=0$, $\mu$
does not develop a peak. Instead, the St{\"u}ckelberg field for these submodels
always decreases monotonically in time. Since this behavior is already captured by the
blue and green examples, we do not demonstrate that explicitly here. Let us only note that
for all these submodels, $\mu$ starts to deviate from its asymptotic value $3(1+w_{\rm m})$
as soon as the Hubble rate falls below the spin-$2$ mass.}.
Below the energy scale, $\mFP$ the St{\"u}ckelberg field becomes small and enters its linear regime.
We have already estimated the energy scale at which the St{\"u}ckelberg field transitions from the linear to the non-linear regime in~\cref{eq:critical-hubble-cosmo}.

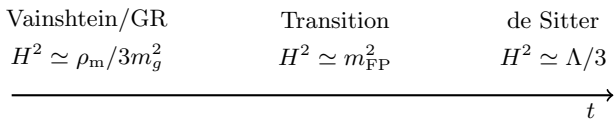
\begin{figure}
\centering
\begin{tikzpicture} 
    \draw[thick] [->] (0,0) -- (8,0);
    \node at (1,1){Vainshtein/GR};
    \node at (1,0.5){$H^2\simeq \rho_{\rm m}/3m_g^2$};
    \node at (4.3,1){Transition};
    \node at (4.3,0.5){$H^2\simeq \mFP^2$};
    \node at (7.2,1){de Sitter};
    \node at (7.2,0.5){$H^2\simeq \Lambda/3$};
    \node at (7.7,-0.2){$t$};
\end{tikzpicture}
\caption{\raggedright Schematic summary of the cosmic evolution in bimetric theory.}
\label{fig:bim-exp-hist}
\end{figure}

In the limit of small $\bar\alpha$ we find the same expression for the critical
Hubble rate as we found with the Vainshtein analogy.
Away from the limit, the critical Hubble rate does not only depend
on $\mFP$ but also on $\bar\alpha$ and $\Lambda$.
This can be seen in~\cref{fig:b0b1b4-stuckelberg} because the inflection
point $y''=0$ and the critical redshift are not exactly aligned.
Our analogy with the local Vainshtein mechanism, hence, gives a rough
estimate of the critical Hubble rate.
The remarkable feature however is, that in the limit
$\bar\alpha\ll1$, the value of the critical Hubble
rate at which the St{\"u}ckelberg field becomes non-linear and background cosmology is screened is set only by $\mFP$ and
becomes independent of $\bar\alpha$ and $\Lambda$.
In \cref{fig:bim-exp-hist}, we show a schematic overview of the different
regimes in bimetric cosmology.

\begin{figure}
\centering
\includegraphics[scale=0.7]{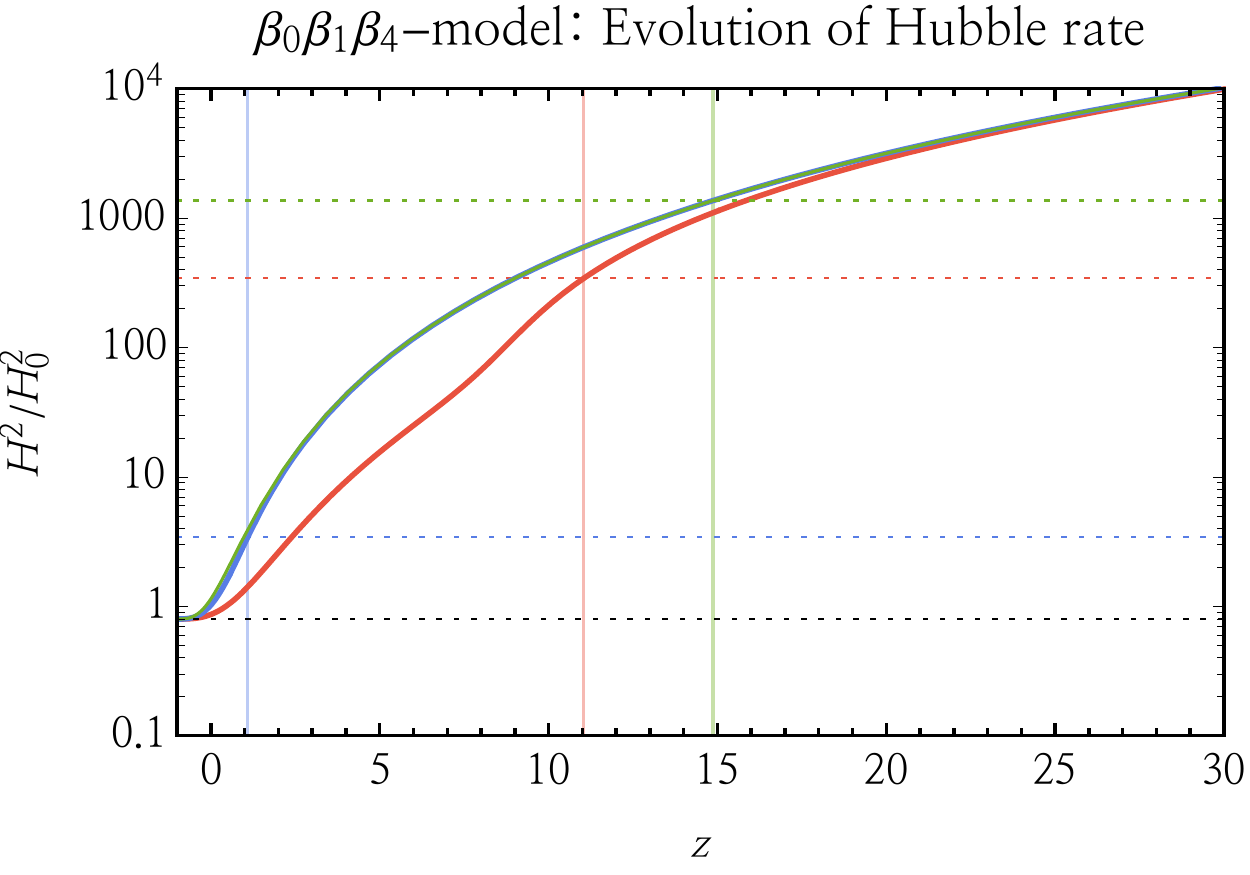}
\caption{\raggedright
The Hubble rate (normalized to $H_0$) as a function of redshift $z$ is shown. 
For all lines we set
$\Lambda=0.7H_0^2$ and $\rho_\mathrm{m,0}/m_g^2=0.3H_0^2$ as exemplary values.
The bimetric parameters are $\mFP=H_0$ and $\bar\alpha=0.5$ (blue line),
$\mFP^2=10H_0^2$ and $\bar\alpha=2$ (red line), and
$\mFP^2=20H_0^2$ and $\bar\alpha=0.01$ (green line).
The vertical lines represent the redshift at which $H=\mFP$, respectively
The horizontal dashed lines indicate $(\mFP/H_0)^2$ (blue, red, green) and $\Lambda/3 H_0^2$
(black).}
\label{fig:b0b1b4-hubble}
\end{figure}

Next, we study the Hubble rate.
In~\cref{fig:b0b1b4-hubble} we show the Hubble rate as function of redshift, for several exemplary values.
Again, the Hubble rate starts to deviate from the energy scale set by
$\rho_{\rm m}/m_g^2$ when $H\sim\mFP$, as indicated by the vertical lines.
However, the deviations are suppressed by $\bar\alpha$.
Therefore, the green ($\bar\alpha=0.01$) and blue ($\bar\alpha=0.5$) line almost coincide.
For the red line, $\bar\alpha=2$ is not small and the transition between the two phases
can be seen directly from the plot. Since the parameters that lead to the red line
imply $\beta_0<0$, the energy contribution from the bimetric potential is negative for
sufficiently large redshift. This implies that the Hubble rate is smaller
than the energy scale set by $\rho_{\rm m}/m_g^2$.

To conclude, background cosmology of bimetric theory resembles GR when the massless and massive mode decouple.
They decouple when $\bar y\ll 1$.
This can be achieved either by adjusting $\alpha\ll1$, i.e. by going to the GR-limit of the theory.
Then, deviations from GR are suppressed at all redshifts.
Alternatively, $\bar y\ll 1$ for $H\gg \mFP$ as we have just shown.
In this case, the energy density due to the non-linear spin-2 interaction is screened and hence negligible above the same energy scale that we encountered in~\cref{sec:Vainshtein-cosmo}.

\subsection{Linear scalar perturbations and Vainshtein}\label{sec:perturbations}

In this section, we connect our previous results to cosmological perturbations around
the FLRW background in bimetric theory.
Their analysis received a lot of attention in the
literature~\cite{Konnig:2014dna,Konnig:2014xva,Konnig:2015lfa,Lagos:2014lca,DeFelice:2014nja,Akrami:2015qga,Aoki:2015xqa}
and here we will summarize and rephrase the conclusions. It was found that the linear
scalar perturbations in the WKB approximation are unstable on subhorizon scales
during early times (for an expansion history on the finite branch).
More precisely, they are stable as long as the dynamical
bound~\cite{Konnig:2015lfa},
\begin{flalign}\label{eq:cosmo-pert-stability}
    y^{\prime\prime} < \frac{y^\prime}{2y}\frac{2\mathcal H y^\prime (y^\prime-3w_m y)-3a^2\rho_{\rm m} y^2 (w_m+1)(2w_m+1)}{a^2\rho_{\rm m} y(w+1)+\mathcal H^2 y^\prime}\,,
\end{flalign}
is satisfied. For models with $\beta_2=\beta_3=0$ this is equivalent to
$y''<0$ \cite{Konnig:2015lfa}.

Hence, the linear perturbations are unstable exactly when the background is screened.
In other words, exactly when the St{\"u}ckelberg field becomes non-linear, linear perturbation theory breaks down.
For the $\beta_0\beta_1$- and $\beta_1\beta_4$-model we already identified the energy scale at which $y''=0$.
For completeness, we also compute at which energy scale the dynamical bound in~\cref{eq:cosmo-pert-stability} is violated for the remaining two parameter models, $\beta_1\beta_2$ and $\beta_1\beta_3$.
The expression for the critical Hubble rate is too long to display it here, but in the limit $\bar\alpha\ll 1$ we find $H_*\simeq\mFP$ (up to an
$\mathcal O(1)$ factor)\footnote{For the
$\beta_1\beta_2$-model, $H_*$ was already computed in Ref.~\cite{Akrami:2015qga}
in the limit $\alpha\ll1$, but not in terms of the physical parameters. They did not interpret their result as the spin-$2$ mass.} as well.

The expansion history and the scalar perturbations are sensitive to the energy scale set by $\mFP$ as we have explicitly demonstrated for all the two-parameter models.
When the Hubble rate is of the order of the spin-$2$ mass, $H\sim \mFP$, the St{\"u}ckelberg field $\mu$ becomes non-linear and the linear perturbations start to grow exponentially.
At the same energy scale, the universe becomes smaller than its own Vainshtein radius.
Combining these results suggests that the Vainshtein mechanism is active also on a time-dependent background like the FLRW and with spatially extended matter sources.
Our result suggests that the scalar perturbations are cured by the Vainshtein mechanism.

Indeed, the stability of scalar perturbations around FLRW background was studied in the literature.
The authors of Ref.~\cite{Aoki:2015xqa} solved the perturbation equations
non-linearly for early times\footnote{In~\cite{Aoki:2015xqa},
spherically symmetric perturbations on subhorizon scales and
on length scales smaller than the Compton wavelength $\mFP^{-1}$ were studied.
Furthermore, the background was assumed to be proportional, $\bar g_{\mu\nu}=y^2\bar f_{\mu\nu}$, with a time-dependent conformal factor $y$.
This is only a solution to the equations of motion when both metrics couple to their own matter sector, which are proportional on-shell.}.
Their analysis identifies the spin-$2$ mass $\mFP$ as the scale at which the
Vainshtein mechanism kicks in and restores GR (for a generic model).
In Ref.~\cite{1910.01651}, on the other hand, the equations of motion were solved
for an inhomogeneous mass distribution non-linearly.
Again, no instabilities were found. Both these results show that
the instabilities are indeed an artifact of the linear approximation and that the Vainshtein
mechanism is active also on a time-dependent background with a spatially extended
source. In a different setting, the traditional Vainshtein mechanism was used to
investigate the effect of early time instabilities on structure formation, see Ref.~\cite{1506.04977}.

\section{Impact on the $H_0$ tension}\label{sec:H0}

Despite the huge success of General Relativity describing gravitational systems on many
different scales to enormous precision and in particular of the $\Lambda$CDM model,
latest data challenge the Standard Model of Cosmology.
Local observations of the Hubble flow are in good agreement with each other and constrain the value of the Hubble
rate today to $h = 0.7324 \pm 0.0174$~\cite{Riess:2016jrr,Mortsell:2018mfj} where
\begin{flalign}
    h = \frac{H_0}{100 \unit{km/s/Mpc}}
\end{flalign}
is the normalized Hubble rate today. 
Recently, a new measurement of the local Hubble rate was performed using Megamasers~\cite{2001.09213}. This observation relies on independent distance determinations and thus has no overlap in systematic errors with previous studies. A Hubble rate value of $h = 0.739 \pm 0.030$ was found, in agreement with previous local measurements.

In the $\Lambda \rm CDM$ model, CMB data from the \textit{Planck} experiment, 
however, favor a value $h = 0.6781 \pm 0.0092$~\cite{Ade:2015xua}. 
This causes a $\sim 3.4\sigma$ tension with the Hubble rate value from local observations.
While local measurements are quite sensitive to
systematics~\cite{Follin:2017ljs,Dhawan:2017ywl,Feeney:2017sgx}, the constraints from
CMB measurements highly depend on the gravitational model~\cite{Ade:2015xua}.
Indeed, CMB data alone favors a Dark Energy component with a
phantom equation of state, $w_\mathrm{DE}<-1$~\cite{Ade:2015rim}.

\subsection{Phantom and negative dark energy}

Several mechanisms were proposed to alleviate the discrepancy in the value
of $H_0$ inferred from late-time and CMB measurements.
One possible direction is to lower the Hubble rate via late-time
modifications of the $\Lambda\rm CDM$ model. This can be achieved by a
dynamical dark energy component that grows as the universe expands (phantom dark energy)
or that changes sign at a certain redshift, i.e. with a negative energy density at sufficiently early
times.
Bimetric theory features both mechanisms depending on the
parameters. 

Models with a phantom equation of state should be treated with caution. Phantom
energy
violates the dominant energy condition~\cite{Hawking:1973uf} and causes a future
spacetime singularity
(Big Rip)~\cite{Caldwell:1999ew,Caldwell:2003vq,Frampton:2002vv,Nesseris:2004uj}.
For simple models that build on a single field (such as quintessence) a phantom equation of state
implies the presence of a low-energy ghost~\cite{Carroll:2003st}.
There are ways to get around these issues.
The Big Rip can be avoided if the equation of state varies in time and approaches
$-1$ sufficiently fast~\cite{Nojiri:2005sx,Nojiri:2005sr,Stefancic:2004kb}.
Ghost condensation can stabilize the vacuum~\cite{ArkaniHamed:2003uy}.

In contrast to these simple realizations,
bimetric theory incorporates an effective phantom equation of state naturally.
The dark energy density $\rho_{\rm DE}$ contained in Friedmann's~\cref{eq:mod-friedmann} 
has the equation of state
\begin{flalign}
    w_{\rm DE}=-1-\frac{\alpha^2y^2}{1+\alpha^2y^2}\frac{(1+w_{\rm m})\rho_{\rm m}}{\rho_{\rm DE}}\frac{m_\mathrm{eff}^2}{m_\mathrm{eff}^2-2H^2}\,.
\end{flalign}
A cosmic expansion history on the finite branch implies $m_\mathrm{eff}^2>2H^2$ and hence
for $\rho_{\rm DE}>0$ the equation of state is phantom,
$w_\mathrm{DE}<-1$~\cite{Konnig:2015lfa}.
At early times the asymptotic value
of $w_{\rm DE}$
is either $-1$ (for models with $\beta_0\ne0$) or $-2-w_{\rm m}$
(for models with $\beta_0=0$).
In the asymptotic future (when $\rho_{\rm m}\rightarrow0$) the equation of state
approaches $w_{\rm DE}\rightarrow-1$ and the effect of the dynamical Dark Energy
reduces to a cosmological constant.
Hence, the Big Rip is avoided in bimetric theory.

Moreover, the effective dark energy component $\rho_{\rm DE}$ can change its sign.
While in the infinite future the dynamical dark energy component approaches the value of the asymptotic cosmological constant, at early times $\rho_{\rm DE}/m_g^2\rightarrow\beta_0$ because $y\rightarrow0$ on the finite branch.
Hence, if $\beta_0<0$ the dark energy component is negative above a certain redshift.
For the simple $\beta_0\beta_1\beta_4$-model this redshift is determined by $y=-\beta_0/(3\beta_1)$ as follows from~\cref{eq:dark-energy-density}.

\begin{figure}
\centering
\includegraphics[scale=0.65]{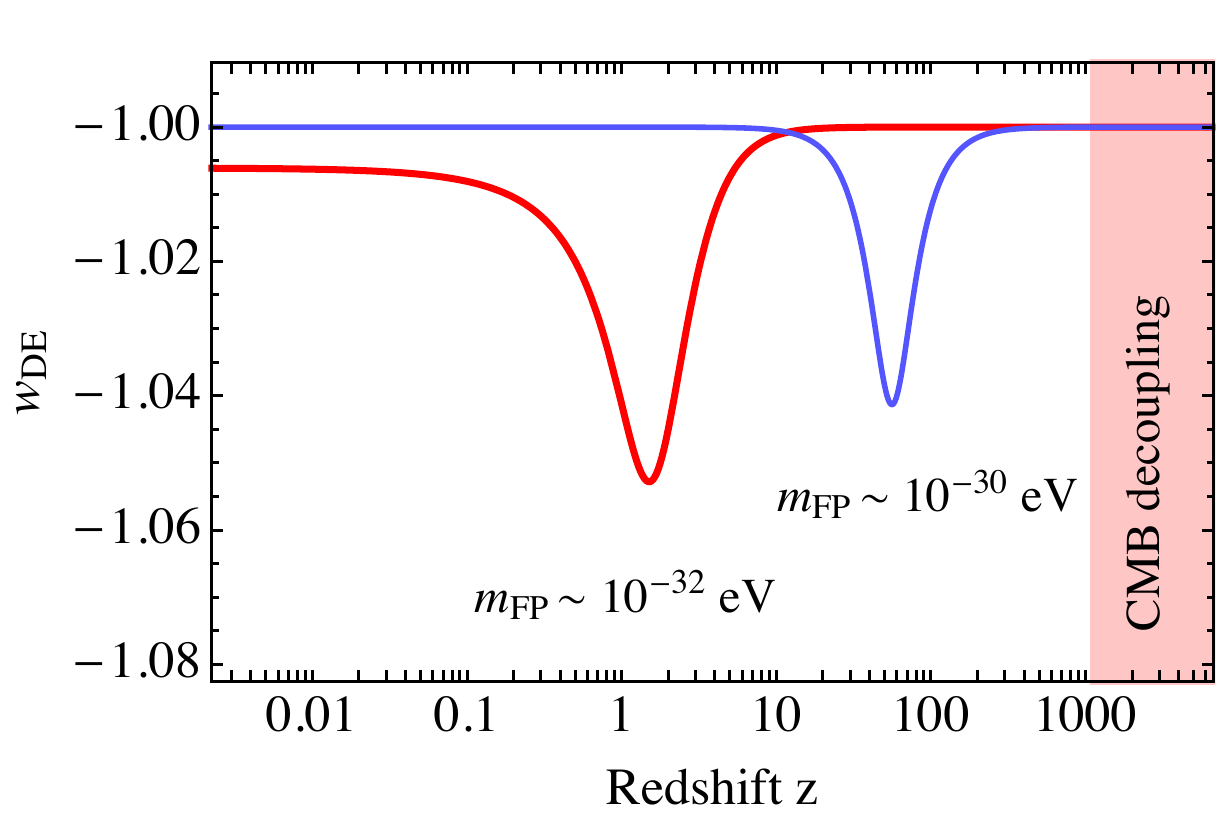}
\caption{\raggedright The equation of state of the effective dark energy component. Two different scales for the spin-$2$ mass are chosen.
The lighter corresponds to the Hubble scale today (red line) and the heaver to the Hubble scale
at redshift $z \simeq 100$ (blue line).}
\label{fig:eosGen}
\end{figure}

In Fig.~\ref{fig:eosGen}, we show the time evolution of the Dark Energy equation
of state for two different graviton masses. 
The phantom era of the cosmic evolution takes place when the size of the universe is
comparable to the wavelength of the graviton, i.e. $H(z) \simeq m_{\rm FP}$ and is related
to the transition from the Vainshtein to the de Sitter regime.

The effective Dark Energy component violates the null energy
condition (NEC)~\cite{Baccetti:2012re} allowing for a phantom equation
of state while the sum of potential energy and matter stress-energy
satisfies the NEC.
In bimetric theory, this does not imply
the presence of a ghost mode due to the nontrivial interactions between the different
modes that give rise to the effective phantom dark energy. On the other hand,
the NEC violation manifests itself in linear perturbation theory as the gradient
instability. However, as we argued earlier, it appears to be an artifact of the calculation and higher-order terms have to be taken into account
due to the Vainshtein mechanism. It would be interesting to study the connection to ghost condensation.

In the following section we perform a fit to data. We restrict ourselves to the study of
the $\beta_0\beta_1\beta_4$-model. 
It is the simplest minimal bimetric model where all three physical parameters
(mixing parameter $\bar\alpha$, spin-$2$ mass $\mFP$, and cosmological constant $\Lambda$)
are independent of each other.
Already the authors of Ref.~\cite{Mortsell:2018mfj} studied
phantom dark energy as a possible resolution to the $H_0$-tension\footnote{Additionally
in~Ref.~\cite{2002.01487}, the inverse distance ladder method
is discussed with the goal of testing the parameter region in bimetric theory, relevant
for the $H_0$ tension.}.
In particular, the bimetric $\beta_0\beta_1$- and $\beta_1\beta_2$-model were used as
concrete models. They conclude that these models are driven into their GR-limits and hence
do not resolve the $H_0$-tension. However, for these two-parameter-models of bimetric theory, 
not all the physical parameters are independent and the setup is too restricted. In particular,
the fact that at small redshifts the equation of state is forced to be $w_{\rm DE} \simeq -1$
enforces a small mixing parameter $\bar{\alpha}\ll1$. This is the GR-limit of these models
and implies an equation of state close to $-1$ at all redshifts.
This is not the case for models where the three physical parameters are free
as we will demonstrate in this section with the $\beta_0\beta_1\beta_4$-model.
We find that solutions exist which feature $w_{\rm DE} \simeq -1$ at small redshifts 
but deviate from that value at intermediate redshifts. 
In~\cref{append:b0b1b4} we collect the equations that we need for the analysis.

\subsection{Parametrization and scanning strategy}

To treat cosmological observables of late and early times on somewhat equal footings, we
do the following. We approximate Friedmann's equation at late times by the Hubble law including the 
deceleration parameter $q=-(1+\dot H/H^2)$ as
\begin{flalign}
H(z) = H_0 + (1+q)H_0\, z\,.
\end{flalign}
Observations of Cepheid variables constrain the Hubble rate parameter today to be $ h = 0.7324 \pm 0.0174$~\cite{Riess:2016jrr}. The quoted value is derived from the combination of four independent Cepheid observables, which we will use to constrain the local Hubble rate. 

Furthermore, we use  observations of Type Ia supernovae to constrain the deceleration parameter.
We consider supernovae with redshifts $z<0.5$ such that the description in terms of the
deceleration parameter is still valid. We find the best-fit value $q= -0.55\pm0.1$ which is
consistent with the results of
Refs.~\cite{1102.3237, astro-ph/9805201,1811.02376}. This approach projects a large number of SN-Ia observables on one coarse-grained parameter. We choose this description, to have a similar number of local and global observables. 
We restrict our analysis to the supernovae with moderate redshifts, as high redshift supernova observations are subject to substantial luminosity uncertainties~\cite{1912.04903}, and might be affected stronger by anisotropy effects~\cite{1808.04597}. 

One crucial point of the present analysis is the following. We assume that the physics that controls the inhomogeneities of the CMB in bimetric theory is identical to GR. 
This assumption is justified since the Vainshtein mechanism ensures that GR is restored at early times at the background level. 
For our analysis, we further assume, that small perturbations around this GR background are well described by the standard CMB perturbation theory.

The essence of the CMB physics, can be well captured by the following
coarse-graining method, suggested in Ref.~\cite{Mortsell:2018mfj}.
At the core of the analysis, there are only three physical observables. Two of which are the
shift parameters based on the comoving angular distance to the last scattering surface 
\begin{align}
 D_A(z^*)= \int_0^{z*} \frac{dz }{H(z)}\, \text{ given that } \Omega_K = 0\,, 
\end{align} 
where $z^*$ is the redshift at which decoupling happens, and the sound horizon
\begin{align}
r_s(z^*)= \int_0^{a*} \frac{c_s da }{a^2H(a)} = \frac{1}{\sqrt{3}}\int_0^{a*} \frac{ da }{a^2H(a)\sqrt{1+ \frac{3 \Omega_\mathrm{b}}{4 \Omega_\gamma}  a}}   \,,
\end{align}  
where $c_s$ is the speed of sound that depends on the energy density of
baryons $\Omega_\mathrm{b}$ and of photons $\Omega_\gamma$.
The physically constrained combinations are 
\begin{itemize}
\item the angular distance normalized to the Hubble horizon at decoupling $\mathcal{R} = \sqrt{\Omega_\mathrm{m,0} }\,H_0 D_A(z^*)$,
\item and the principle multipole number $ l_A = \pi \frac{ D_A(z^*) }{r_s(z^*)}$.
\end{itemize}
Here, $\Omega_\mathrm{m,0}=\rho_{\rm m}/(3m_g^2 H_0^2)$
is the energy density today of non-relativistic matter. 
The third parameter is the energy density of baryons at decoupling $\Omega_\mathrm{b}h^2$.
We use the CMB compressed likelihood~\cite{Ade:2015xua} with values
$(\mathcal R, l_A, \Omega_\mathrm{b}h^2)=(1.7382,301.63,0.02262)$, errors $(0.0088,0.15,0.00029)$
and the covariance matrix 
\begin{flalign}
    D_\mathrm{CMB} =
    \left(\begin{matrix}
        1.0 & 0.64 & -0.75 \\
        0.64 & 1.0 & -0.55 \\
        -0.75 & -0.55 & 1.0
    \end{matrix}\right)\,.
\end{flalign}
Note that in our analysis, we modify the analysis of Ref.~\cite{Mortsell:2018mfj} by adding a weight factor in the $\chi^2$ function. The weight factor $w_p$ multiplies the contribution of $l_A$ to the $\chi^2$ funciton, and thus takes into account the peak multiplicity in the CMB spectrum. We take $w_p = 7$, as it corresponds to the number of peaks, well resolved by the Planck experiment. This approach approximately models the statistical weight of the Planck data in comparison to the local observables. 

The same physical scale of the CMB perturbations is imprinted in the matter power spectrum as
the baryon-accoustic-oscillations (BAOs), and is accessible to us in the data of several surveys,
measuring galaxy distributions at different redshifts.
The useful oblique parameter relevant for the computation of the matter distribution observable
is the ratio of the sound horizon $r_s(z_d)$ and the spherical average of the angular scale
and the redshift separation $d_z = r_s(z_d)/D_V(z)$, where
\begin{align}
D_V(z) = \left( D_A(z)^2 \frac{z}{H(z)}\right)^{1/3}\,,
\end{align}
and $z_d$ is the drag epoch~\cite{astro-ph/9709112}, the redshift at which the baryons are
released from the Compton drag of the photons. 
We consider four experimental values of $d_z$ at different effective redshifts, reported by 6dFGS~\cite{1106.3366}: $d_z = 0.34 \pm 0.02$ at $z_{\rm eff} = 0.106$, SDSS~\cite{1312.4877}: $d_z = 0.22 \pm 0.01$ at $z_{\rm eff} = 0.15$, BOSS~\cite{1706.03630}: $d_z = 0.118 \pm 0.002$ at $z_{\rm eff} = 0.32$ and $d_z = 0.072 \pm 0.001$ at $z_{\rm eff} = 0.57$.

Given this experimental input (i.e. from Cepheids, SNIa, BAOs, and CMB),
we perform a $\chi^2$ analysis. As a reference,
we scan the $\Lambda \rm CDM$ model in the region $h = 0.65 - 0.75$,
$\Omega_{\Lambda} = 0.6 - 0.8$, $\Omega_\mathrm{b ,0}= 0 - 0.1$,
and $\Omega_\mathrm{\gamma,0}= 0 - 10^{-2}$. The value of $\Omega_\mathrm{m,0}$
is fixed by the flatness condition. For the $\beta_0\beta_1\beta_4$-model we scan
over the same region and additionally have the free parameters $\bar\alpha= 0 - 1$
and $\Omega_{\mFP}  =\mFP^2/(3H_0^2)$  in the range
$\Omega_{\mFP} = 0 - 50$. We perform at first a linear grid scan and refine the $\chi^2$
fit by a Metropolis-Hastings method.
\begin{figure}
\centering
\includegraphics[scale=0.65]{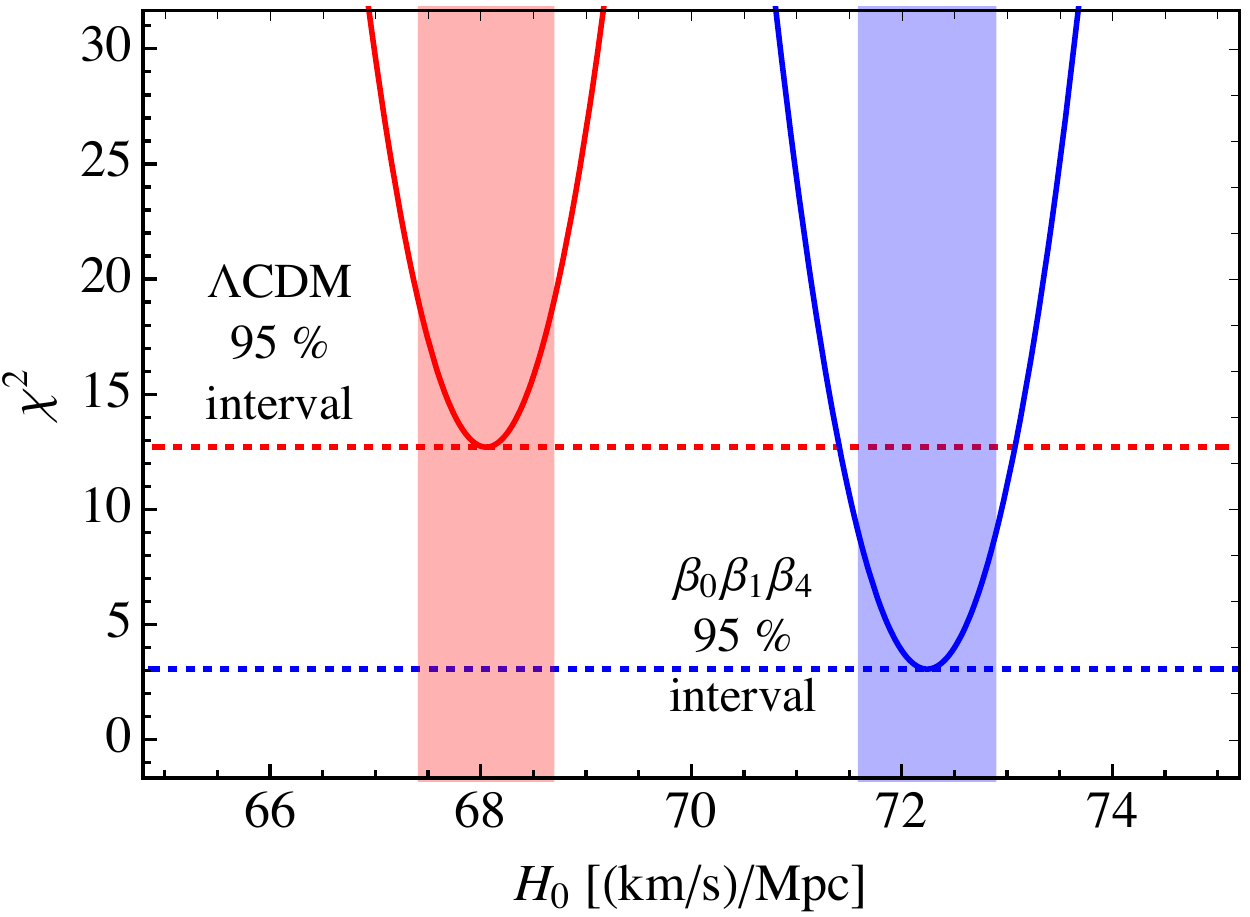}
\caption{\raggedright The $\chi^2$ functions and $95\%$ confidence intervals of $H_0$,
around the best-fit points of the $\Lambda \rm CDM$ model
and the $\beta_0\beta_1\beta_4$-realization of bimetric theory.
The fit improvement is substantial, with $\Delta \chi^2/\text{dof} \approx 5$.}
\label{fig:H0fit}
\end{figure}
%

\subsection{Numerical results}

In Tab.~\ref{tab:results}, we show the best-fit values and one sigma intervals of the $\Lambda \rm CDM$ and the $\beta_0 \beta_1 \beta_4$ model. As expected, the $\Lambda \rm CDM$ model fit is poor.
The best-fit value of $\chi^2\approx 14$ with seven degrees of freedom suggests a $\sim 3 \sigma$ tension of
the global fit. The error intervals are derived from projections on the one dimensional
subspaces of the likelihood function.

In contrast to this, when the fit is performed in the $\beta_0 \beta_1 \beta_4$ model, the fit
is improved and $\chi^2 \approx 3$. Given that we have only two additional fit-parameters, resulting in five degrees of freedom, this corresponds to an excellent fit value, indicating that all
observables are within the $1 \sigma$ error range. Another measure of the improvement
is the $\Delta \chi^2/\rm dof$, which in this case is $\sim 5$. We conclude that the given
data set favors the $\beta_0 \beta_1 \beta_4$ model. 

\begin{table}[t]
\centering
{\def\arraystretch{1.3}
\begin{tabular}{ l | l l }
\hline\hline
& $\Lambda$CDM & $\beta_0\beta_1\beta_4$\\
\hline
$h\times 100$ & $67.9 \substack{+ 0.4 \\ -0.4} $ & $72.3 \substack{+ 0.4 \\ -0.4}$\\
$\bar{ \alpha}$ & -- & $0.13\substack{+0.02 \\ -0.03}$\\
$\mFP/H_0$ & -- & $0.59 \substack{+0.17 \\ -0.13}$\\
$\Omega_{\Lambda }$ & $0.686 \substack{+0.001 \\ -0.001}$ & $0.707 \substack{+0.001 \\ -0.002}$\\
$\Omega_{\rm b,0}$ & $0.0488 \substack{+0.0004 \\ -0.0004}$ & $0.0430 \substack{+0.0004\\-0.0004}$\\
\hline\hline
\end{tabular}}
\caption{\label{tab:results} \raggedright The best-fit parameter values of the $\Lambda \rm CDM$ and the $\beta_0\beta_1\beta_4$ models.
In both cases $\Omega_\gamma$ is irrelevant for late time observations.}
\end{table}

In Fig.~\ref{fig:H0fit}, we show the $\chi^2$ as a function of $H_0$ for the
$\Lambda \rm CDM$ and the $\beta_0 \beta_1 \beta_4$-model. We find, that the alternative
time evolution of the Hubble rate in bimetric theory can accommodate the CMB observables,
and a larger $H_0$ value today, than the $\Lambda \rm CDM$ scenario, while being consistent
with the BAO observations.
The favored Fierz-Pauli mass in the considered best-fit interval
is $\mFP \approx \left( 4 \cdot  10^{-33} - 7 \cdot 10^{-33}\right)\, \rm eV$,
which is consistent with cluster lensing~\cite{Platscher:2018voh} and other
constraints~\cite{Luben:2018ekw}.

\subsection{Discussion of results}

\begin{figure}
\centering
\includegraphics[scale=0.65]{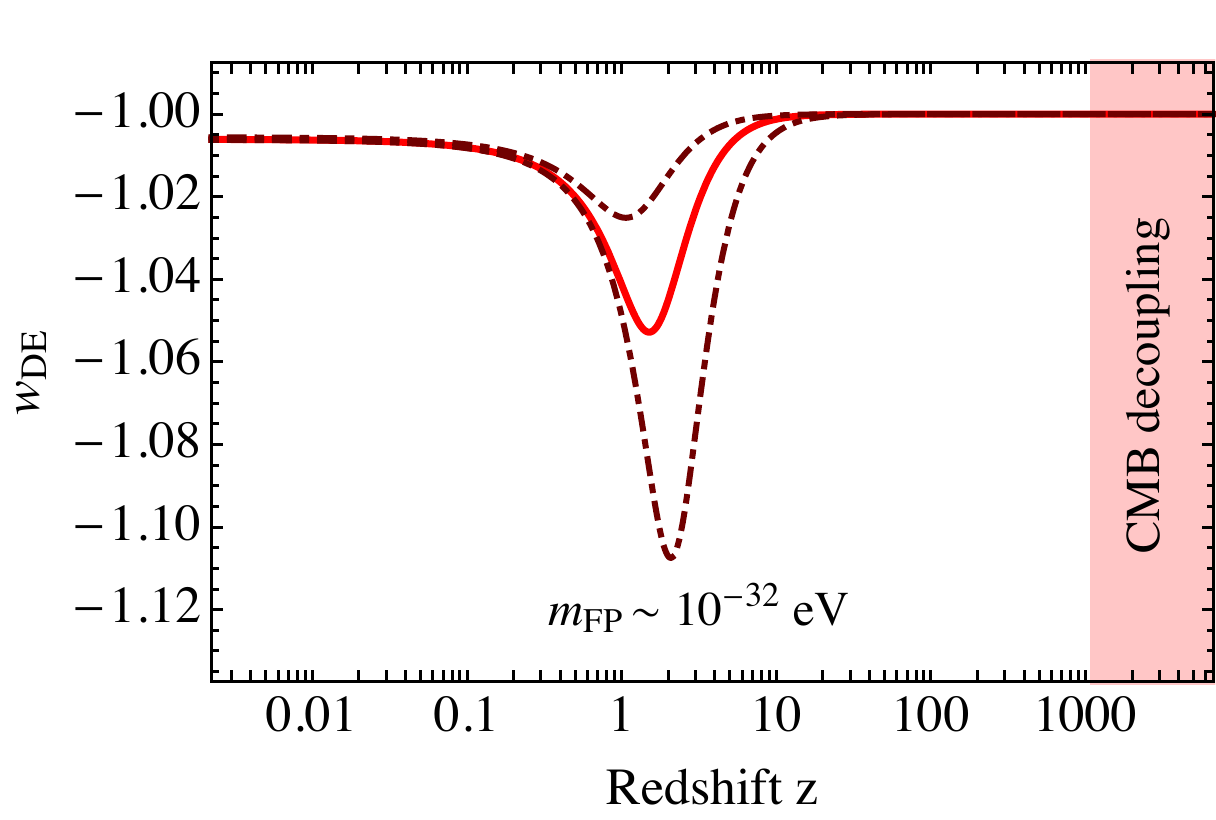}
\caption{\raggedright The equation of state of the effective dark energy component. The minimum of the $\chi^2$ function in the bimetric model corresponds to the red line, see \cref{tab:results}. The dashed lines indicate the best-fit parameter intervals.}
\label{fig:eos}
\end{figure}

Now we discuss the results of the statistical analysis and underlying data sets.
First note that the parameters at the best fit point imply $\beta_0>0$ and hence
$\rho_\mathrm{DE}>0$, cf.~\cref{eq:beta0}. Thus dark energy is always phantom and positive.
In~\cref{fig:eos}, we show the equation of state of the effective dark energy component in
the $\beta_0\beta_1\beta_4$-model at the minimum of the $\chi^2$ function (solid line)
and the $1\sigma$ intervals (dashed lines).
The values are in agreement with current experimental bounds~\cite{1303.4353}.

The phantom behavior is most pronounced in the redshifts interval between $z\sim 1$ and
$z\sim 10$. This is when the Hubble rate is of the order of the spin-$2$ mass.
The general behavior is the following.
The spin-$2$ mass controls at which redshifts the equation of state significantly deviates
from $-1$, while the mixing parameter $\bar\alpha$ controls the magnitude of the deviation.
To be precise,
it is the value of $\beta_0$ that controls the deviation. If its value is close to zero, the
equation of state significantly deviates from $-1$. On the other hand, if $\beta_0$ is positive
and far away from zero, the equation of state is close to $-1$ at all times.
Note that in order to achieve a value $\beta_0$ close to zero requires a non-zero mixing
parameter $\bar\alpha$, cf.~\cref{eq:beta0}. In the GR-limit $\bar\alpha\ll1$ the phantom
era is absent.
The freedom to allow for large spin-$2$ masses and thus shifting the phantom behavior to
larger $z$ while keeping $\bar\alpha$ finite to yield a significant phantom era, is not
possible in the more restricted two parameter models~\cite{Mortsell:2018mfj}.

The phantom behavior of the dynamical dark energy component lowers the Hubble rate
for redshifts where it is still non-negligible, i.e. most importantly for
$z\sim\mathcal O(1)$, as compared to a cosmological constant.
The lowered Hubble rate increases the value of the integral in $D_A(z^*)$ which is compensated
by a larger value of $H_0$ to arrive at the same value of $D_A(z^*)$ as desired.
Since the value of $r_s(z^*)$ remains unchanged by late-time modifications, the value of $l_A$
remains unchanged. However, within $\mathcal R$ the larger value of $H_0$ must be compensated
by a smaller value of $\Omega_\mathrm{m,0}$ and consequently a larger
value of $\Omega_\Lambda$.
This is indeed the case (see \cref{tab:results})\footnote{We
are thankful to E. M{\"o}rtsell for making this point clear.}.

\begin{figure}
\centering
\includegraphics[scale=0.9]{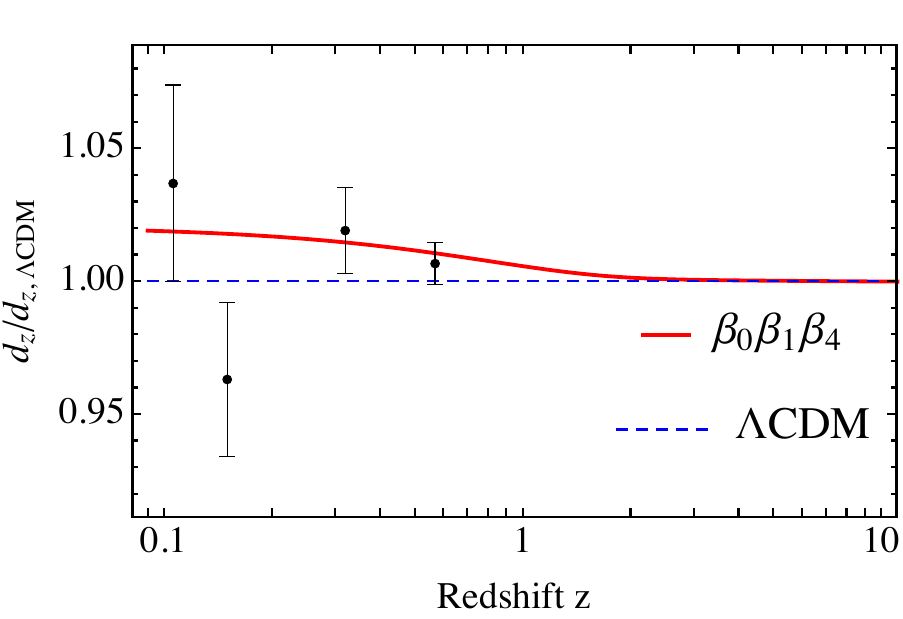}
\caption{\raggedright The ratio of the baryon-acoustic-oscillation parameters normalized to the $\Lambda \rm CDM$ prediction $d_z/d_{z, \,\Lambda \rm CDM}$ in the $\beta_0 \beta_1 \beta_4$-model and the  $\Lambda \rm CDM$ as a function of the cosmological redshift. Superposed are four measurement points at different redshifts.}
\label{fig:BAOfit}
\end{figure}

For the $\Lambda\rm CDM$-model a smaller value of $\Omega_\mathrm{m,0}$ is disfavored
by local measurements as well as the CMB because the matter-dark energy equality
is shifted to higher redshifts, also affecting the value of $l_A$.
To see that bimetric cosmology with a smaller $\Omega_\mathrm{m,0}$
is compatible with local experiments,
we compare the prediction of the $\beta_0\beta_1\beta_4$-model 
(red solid line) to the
four measurements of BAO observables in~\cref{fig:BAOfit}.
For a better resolution, we normalized the plot to
the $\Lambda\rm CDM$ prediction (dashed line).
We can also see that, given the current experimental precision,
BAOs can not distinguish the predictions of the bimetric
$\beta_0\beta_1\beta_4$-model from predictions of
$\Lambda \rm CDM$ at their best-fit points.

However, baryon-acoustic-oscillations might provide a test of bimetric cosmology
at the best-fit point found in this section.
In near future the DESI instrument~\cite{1308.0847,1907.10688} will provide a new dataset of
BAO observations at multiple redshifts and even more advanced experiments will push to
larger redshifts~\cite{1907.11171}. A big advantage of DESI will be, that BAO data will be
obtained at multiple redshifts with the same instrument, thus avoiding the problem of different
systematic errors among the instruments. With the newly collected data, this question
should be re-examined. In particular with an estimated accuracy improvement by a factor
of two, as predicted in Ref.~\cite{1611.00036}, the $\beta_0\beta_1\beta_4$-model at
its best-fit point will be distinguishable from $\Lambda \rm CDM$.

Moreover, measurements of supernovae at high redshifts might allow to distinguish
the bimetric $\beta_0\beta_1\beta_4$-model from the $\Lambda\rm CDM$ model if the
$H_0$-tension should be alleviated within bimetric theory solely due to the phantom phase.
Although we only included supernovae with redshift $z<0.5$ in our analysis,
our best-fit point is consistent also with supernovae measurements at higher
redshifts~\cite{Luben:2020xll} given the current experimental precision~\cite{1912.04903,1808.04597}.

When discussing the significance of the Hubble tension, a word of caution is in order. So far we have taken the local determinations of the Hubble rate at phase value with the quoted uncertainties, which are reported to be at the $2-3\%$ level. However, the distance determination with the Cepheid observables is known to be subject to systematic errors, which are hard to control, see for example Ref.~\cite{astro-ph/9703059}. 

One important effect is the so-called blending effect and is based on the fact that the spatial resolution of our instruments gets worse with distance. Thus observations of Cepheids, that are further away are more likely to pick up light from unresolved background sources. This leads to systematically larger luminosities for more distant objects. This effect tends to increase the reconstructed local Hubble rate, which is consistent with the sign of the observed discrepancy. Taking the blending effect into account would increase the uncertainty to the  $\sim 5\%$ level~\cite{Macri:2006wm,1103.0549}. Even though this would not resolve the Hubble tension, without proper control of the systematic error, we can not make strong statements about the true statistical significance of this anomaly.

\section{Summary and discussion}

In the first part of this paper, we discussed the signatures of the Vainshtein mechanism in
cosmology.
We find that the universe becomes larger than its own Vainshtein radius at the critical Hubble rate
\begin{equation}    
    H_{\rm c}=\mFP\,.
\end{equation}
For earlier times, each Hubble patch lies within its own Vainshtein regime and hence is screened.
We find the same scale appearing on the level of Friedmann's equation and linear perturbations.
In particular, non-linear massless and massive modes decouple at early times, i.e. for $H\gg \mFP$.
Further, the St{\"u}ckelberg field becomes large and hence non-linear above the same scale.
Finally, linear perturbation around the FLRW background are unstable and grow exponentially when $H\gg\mFP$.
We conclude that the early universe is screened by the cosmological analogue of the Vainshtein mechanism.
When studying perturbative phenomena in the early universe, non-linearities have to be taken into account.

We assumed that these non-linearities are such that the massive mode becomes strongly coupled
and can be integrated out to restore GR, as demonstrated in~\cite{Aoki:2015xqa}.
This allows us to study perturbative phenomena, such as the formation of the CMB.

Under this assumption (that CMB physics is the same as in GR due to the Vainshtein mechanism), we perform a global fit to the CMB observables and the local measurements on Cepheid variables, SNIa observations of the cosmic acceleration and baryon-accoustic-oscillations. We find that due to the phantom equation of state of the effective dark energy component in the redshift interval between $z\sim1$ and $z\sim10$ the reported tension between the local and CMB determination of the Hubble scale can be resolved. 

We discussed some potential shortcomings of our statistical analysis. In particular
the blending effect in Cepheid measurements and the high-redshift supernovae should
be taken into account. However, we do not expect these to dramatically change
our result as we discussed.
Instead, if the $H_0$-tension is real and statistically significant,
bimetric theory appears as an experimentally favored, consistent, and theoretically
well-motivated alternative to GR.
Whether data eventually favors bimetric theory also over other gravitational theories,
remains an open question.

We show that current data from BAOs can not distinguish bimetric cosmology
from $\Lambda \rm CDM$. However, in the near future, a strong improvement
in sensitivity and systematic uncertainty will likely change this situation.
Under the requirement that the $H_0$-tension is alleviated solely due to the bimetric
phantom dark energy, the $\beta_0\beta_1\beta_4$-model will be distinguishable from the
$\Lambda\rm CDM$-model in the near future.

A next step would be a global fit to all current data.
In particular, instead of the coarse-grained methods used in our analysis
the whole data sets should be used for studying the statistical significance
of the tension and its alleviation within bimetric theory.
Another direction is a non-linear study of the cosmological perturbations
and of the Vainshtein mechanism.
Within an entirely bimetric framework, the constraints from the CMB should be derived
to explicitly check our assumption a posteriori.
Also, cosmic structure formation should be addressed entirely within bimetric theory
as it probes redshifts where the phantom era occurs.

\vspace{5pt}
\subsubsection*{Acknowledgments} 
We would like to thank Christopher Hirata, Florian Niedermann, Moritz Platscher, Anna Porredon, Fiorenzo Vincenzo, Bei Zhou and especially Edvard M{\"o}rtsell for valuable discussions  and helpful comments on the manuscript. This work is supported by a grant from the Max-Planck-Society. J.S. is largely supported by a Feodor Lynen Fellowship from the Alexander von Humboldt foundation.
The work of M.L. is supported by a grant from the Max Planck Society.

\appendix\section*{Appendix}

\section{Explicit expressions for the $\beta_0\beta_1\beta_4$-model}
\label{append:b0b1b4}
Throughout the paper we used the $\beta_0\beta_1\beta_4$-model that is defined by
setting $\beta_2=\beta_3=0$ to provide an explicit example.
In this appendix, we report the exact expressions and discuss some features of
the model. For details, we refer to~\cite{Luben:2020xll}.

First, let us find the relation between the interaction and physical parameters.
The background~\cref{eq:prop-bkg-consistency} is a cubic polynomial in $c$ and gives
rise to up the three real-valued roots.
Each root describes a vacuum of the $\beta_0\beta_1\beta_4$-model with different
spin-$2$ mass, mixing angle and cosmological constant.
However as discussed in Ref.~\cite{Luben:2020xll}, only one of the vacua is physical.
Therefore, the vacuum~\cref{eq:FP-mass,eq:prop-bkg-consistency} imply the following
unique relation between the interaction parameters and the physical parameters,
\begin{subequations}\label{eq:b0-phys-parameter-b0b1b4}
\begin{flalign}
    \beta_0 & = \frac{ -3 \bar\alpha^2 \mFP^2 + (1+\bar\alpha^2 ) \Lambda }{1+\bar\alpha^2}\label{eq:beta0}\\
    \alpha^{-1} \beta_1 & = \frac{\bar\alpha}{1+\bar\alpha^2} \mFP^2\\
    \alpha^{-4} \beta_4 & = \frac{-\mFP^2 + (1+\bar\alpha^2) \Lambda }{\bar\alpha^2 (1+\bar\alpha^2)}\,.
\end{flalign}
\end{subequations}
The physical parameters are not completely free but have to satisfy the Higuchi bound,
$\mFP^2>3\Lambda/2$, to ensure unitarity~\cite{Higuchi:1986py}.
In the following, we still express all equations in terms of the interaction parameters
$\beta_n$ for brevity but they should be understood as being functions of the physical
parameters. Furthermore, we rescale $\bar y=\alpha y$ for brevity.

Setting $\beta_2=\beta_3=0$, \cref{eq:quartic-pol-y} reduces to
\begin{flalign}\label{eq:y-pol-b014}
    \bar\beta_4\bar y^3-3\bar\beta_1\bar y^2-\left(\beta_0+\frac{\rho_{\rm m}}{m_g^2}\right)\bar y + \bar\beta_1=0\,,
\end{flalign}
where $\bar\beta_n=\alpha^{-n}\beta_n$ for brevity.
This polynomial has up to three real-valued roots that yield $y$ as a function of $\rho_{\rm m}$.
For a given set of parameters, only one of these solutions corresponds to the
finite branch. Since the expressions are quite lengthy and not enlightening, we do not show them
explicitly here. The finite branch solution must satisfy $0\leq\bar y\leq\bar\alpha$ which allows
picking the finite branch solution numerically.
Hence, we can express
\begin{flalign}
    \mu=\frac{y'}{y}
\end{flalign}
either analytically (but lengthy) or numerically as a function of matter-energy
density $\rho_{\rm m}$ and consequently as a function of
redshift $z$ only. We used that for producing the exemplary plots
in~\cref{fig:b0b1b4-stuckelberg}.

Next, we want to find the Hubble rate as a function of redshift $z$ only. For
the $\beta_0\beta_1\beta_4$-model the Hubble rate reads
\begin{flalign}
    3 H^2 = \beta_0+3\bar\beta_1 \bar y +\frac{\rho_{\rm m}}{m_g^2}=\frac{\bar \beta_1}{\bar y}+\bar\beta_4 \bar y^2\,,
\end{flalign}
where the interaction parameters $\bar\beta_n$ are understood as functions of the physical
parameters and $\bar y$ as the finite branch solution to~\cref{eq:y-pol-b014}
(either analytically or numerically).
We used the result for drawing the plot in~\cref{fig:b0b1b4-hubble} and for the
data analysis in~\cref{sec:H0}.

Let us collect some more details for the data analysis with the $\beta_0\beta_1\beta_4$-model.
Evaluating Friedmann's equation today yields a relation among the parameters of the
model,
\begin{flalign}
    3 H_0^2 = \beta_0+3\bar\beta_1 \bar y_0 +\frac{\rho_\mathrm{m,0}}{m_g^2}\,,
\end{flalign}
where the subscript $0$ indicates the value of the quantity at present time.
We used this relation to eliminate $\Omega_{m,0}$ in terms of the other parameters.

In the data analysis, we constrained the deceleration parameter $q$ that is derived
from Friedmann's equation.
We can express the definition more explicit in terms of bimetric parameters as
\begin{flalign}\label{eq:bim-deceleration-param}
    1+q=-\frac{\dot H}{H^2}=\frac{1}{2H^2}\frac{\dd H^2}{\dd\bar y}\bar y'\,,
\end{flalign}
where $\bar y'=\dot {\bar y}/H$ is given by~\cref{eq:y-prime} and
\begin{flalign}
    \frac{\dd H^2}{\dd\bar y}=-\frac{\bar\beta_1}{3\bar y^2}+\frac{2}{3}\bar \beta_4 \bar y\,.
\end{flalign}
Again, with $\bar y$ understood as the finite branch solution, $q$ is a function of redshift only.
For the data analysis, we used the constraints on $q_0$ (i.e. at $z=0$) to find the
favored values of the physical parameters. 
Note that~\cref{eq:bim-deceleration-param} holds for any bimetric (sub)model.


\footnotesize
\bibliographystyle{abbrv}


\end{document}